\documentclass[%
aps,
sd,%
amsmath,amssymb,
preprint,%
reprint,%
]{revtex4-1}

\usepackage{graphicx}% Include figure files
\usepackage{bm}% bold math
\begin{document}
	
	\preprint{AIP} %{/123-QED}
	
	\title%[Sample title]
	{Exact closed probability-free kinetic equation for system of classical particles with retarded interactions: towards microscopic foundation of kinetics}
	
	\author{A. Yu.~Zakharov}
	\altaffiliation[]{Physics Department, Yaroslav-the-Wise Novgorod State University, Veliky Novgorod, 173003, Russia.}
	
\email{Anatoly.Zakharov@novsu.ru;\\ A.Yu.Zakharov@gmail.com}	

\begin{abstract}
The exact closed equation of motion for microscopic distribution function of classical many-body system with account of interactions retardation between particles is derived. It is shown that interactions retardation leads to irreversible behaviour of many-body systems.  

\end{abstract}
\maketitle

\section{Introduction}
Despite enormous advances of statistical mechanics in description of equilibrium properties and transport processes in condensed matter, the problem of \textit{non-contradictory} microscopic foundation of both thermodynamics and kinetics remains unsolved. Currently, conventional microscopic foundation of thermodynamics and kinetics is based on the union of the probabilistic approach and classical mechanics. Combining deterministic classical mechanics with the concept of probability in the absence of a real source of randomization leads to self-contradictory theory. 

For this reason, from time to time there are heated discussions on the use of probabilistic concepts in the classical theory of many-body systems (see., e.g., the discussion of this issue at the round table during 20th IUPAP International Conference on Statistical Physics~\cite{Leb} ).

The first attempts to use the laws of classical mechanics, coupled with the ``molecular chaos'' (Sto{\ss}zahlansatz) probabilistic hypothesis by Boltzmann at the end of the XIX century, were immediately subjected to the very essential criticism from Loschmidt and Zermelo. They proved that Boltzmann's probabilistic assumptions contradict the exact theorems of classical mechanics~\cite{Kac1,Kac,Strien}. 

Thus, the introduction of the probability concept into classical many-body dynamics allows to describe the irreversible phenomena, but this union of classical mechanics and probability theory is internally inconsistent.

In this connection it should be noted an exactly solvable discrete dynamical model (the ring model), proposed by Mark Kac~\cite {Kac1, Kac}. This model has the properties of reversibility and recurrence~\cite {Gottwald, Kozlov}, but loses these properties with the introduction of very plausible probabilistic hypotheses. Therefore, the probability approach to foundation of thermodynamics and kinetics is not satisfactory. Exact theorems of classical mechanics are at variance not only with the probabilistic description of macroscopic systems, but also with the thermodynamic behavior of systems.

All these facts lead to the conclusion that microscopic foundation of thermodynamics in terms of classical mechanics is impossible. One of the factors that could cause  irreversible behavior systems is retardation of the interactions between the particles~\cite{ZZ}. Since the interactions delay is a relativistic effect, the description of the system evolution should be based on the theory of relativity. According to Currie's ``The No Interaction Theorem''\cite{Currie2,Currie,Trump}, a Hamiltonian description of a system of \textit{interacting particles} in terms of their coordinates~$q_s$ and momenta~$p_s$ is inconsistent with the principle of relativistic unvariance. In other words, the Hamiltonian relativistic description of the system is possible in the case of non-interacting particles only (ideal relativistic gas).

There are many works devoted to relativistic kinetic theory~\cite{Trump,Liboff,Groot,Hakim,Cerc}. Most of them contain relativistic generalization of the collision integral in the Boltzmann equation, the transformation properties of distribution functions, etc. However, the basis of all these works is the concept of probability. 

This paper aimed on derivation and analysis of the exact closed relativistic equation of motion for the microscopic (non-probabilistic) distribution function of a classical (non-quantum) system of interacting particles.

\section{Derivation of the basic equation}

Let us define the microscopic distribution function in 6-dimensional  coordinate-velocity space~$(\mathbf{r,v})$ 
\begin{equation}\label{f(r,v,t)}
	\begin{array}{c}
		{\displaystyle 	f(\mathbf{r},\mathbf{v},t) = \sum_{s=1}^N\ \delta\left(\mathbf{r} - \mathbf{R}_s(t) \right) \, \delta\left(\mathbf{v} - \dot{\mathbf{R}}_s(t) \right)  }\\
		{\displaystyle = \iint\, \frac{d\mathbf{k}}{\left( 2\pi\right)^3 }\, \frac{d\mathbf{q}}{\left( 2\pi\right)^3 } \,e^{i\,\mathbf{k}\,\mathbf{r}} \,e^{i\,\mathbf{q}\,\mathbf{v}} \ \left[ \sum_s \,e^{- i\, \mathbf{k}\, \mathbf{R}_s \left(t \right) } \,e^{-i\,\mathbf{q}\, \dot{\mathbf{R}}_s \left(t \right) } \right]  }.
	\end{array}
\end{equation}
The calculation of the sums of type $ \sum_s \psi \left( \mathbf{R}_s(t), \dot{\mathbf{R}}_s(t) \right)$, where $\psi \left( \mathbf{R}_s(t), \dot{\mathbf{R}}_s(t) \right)$ is any ``one-particle'' function, will perform according to the rule
\begin{equation}\label{mean-psi}
\begin{array}{c}
{\displaystyle \sum_s\, \psi \left( \mathbf{R}_s(t), \dot{\mathbf{R}}_s(t) \right)}\\ {\displaystyle = \sum_s\ \iint d\mathbf{r}\, d\mathbf{v}\, \psi \left(\mathbf{r}, \mathbf{v} \right) \, \delta\left(\mathbf{r} - \mathbf{R}_s(t) \right) \, \delta\left(\mathbf{v} - \dot{\mathbf{R}}_s(t) \right) }\\
{\displaystyle = \iint d\mathbf{r}\, d\mathbf{v}\ f(\mathbf{r},\mathbf{v},t)\, \psi(\mathbf{r},\mathbf{v}).}
\end{array}
\end{equation}

After differentiating the distribution function~(\ref{f(r,v,t)}) with respect to time, we have
\begin{equation}\label{dot1+2-f}
\frac{\partial f(\mathbf{r},\mathbf{v},t)}{\partial t} = \left( \frac{\partial f(\mathbf{r},\mathbf{v},t)}{\partial t}\right) _{1} + \left( \frac{\partial f(\mathbf{r},\mathbf{v},t)}{\partial t}\right) _{2},
\end{equation}
where
\begin{equation}\label{dot1-f}
\begin{array}{c}
{\displaystyle \left( \frac{\partial f(\mathbf{r},\mathbf{v},t)}{\partial t}\right) _{1} = \iint\, \frac{d\mathbf{k}}{\left( 2\pi\right)^3 }\, \frac{d\mathbf{q}}{\left( 2\pi\right)^3 } \,e^{i\,\mathbf{k}\,\mathbf{r}} \,e^{i\,\mathbf{q}\,\mathbf{v}} }\\ \\
{\displaystyle \times \sum_s \,e^{- i\, \mathbf{k}\, \mathbf{R}_s \left(t \right) } \,e^{-i\,\mathbf{q}\, \dot{\mathbf{R}}_s \left(t \right) } \left[- i\, \mathbf{k}\, \dot{\mathbf{R}}_s \left(t \right)  \right],}
\end{array}
\end{equation}
and
\begin{equation}\label{dot2-f}
\begin{array}{c}
{\displaystyle \left( \frac{\partial f(\mathbf{r},\mathbf{v},t)}{\partial t}\right) _{2} = \iint\, \frac{d\mathbf{k}}{\left( 2\pi\right)^3 }\, \frac{d\mathbf{q}}{\left( 2\pi\right)^3 } \,e^{i\,\mathbf{k}\,\mathbf{r}} \,e^{i\,\mathbf{q}\,\mathbf{v}} }\\ \\
{\displaystyle \times  \sum_s \,e^{- i\, \mathbf{k}\, \mathbf{R}_s \left(t \right) } \,e^{-i\,\mathbf{q}\, \dot{\mathbf{R}}_s \left(t \right) } \left[- i\, \mathbf{q}\, \ddot{\mathbf{R}}_s \left(t \right)  \right].}
\end{array}
\end{equation}

Let us express both terms $\left( \frac{\partial f(\mathbf{r},\mathbf{v},t)}{\partial t}\right)_{1} $ and $\left( \frac{\partial f(\mathbf{r},\mathbf{v},t)}{\partial t}\right)_{2} $ via the microscopic distribution function~$f(\mathbf{r},\mathbf{v},t)$.

\subsection{Calculation of  $\left( \frac{\partial f(\mathbf{r},\mathbf{v},t)}{\partial t}\right) _{1}$ and $\left( \frac{\partial f(\mathbf{r},\mathbf{v},t)}{\partial t}\right) _{2}$}

First we transform the right-hand side of~(\ref{dot1-f}), using the rule~(\ref{mean-psi}) and the Fourier representation for delta-function. As a result we obtain  
\begin{equation}\label{dot-f1}
\begin{array}{c}
{\displaystyle \left( \frac{\partial f(\mathbf{r},\mathbf{v},t)}{\partial t}\right) _{1} =  \iint\, \frac{d\mathbf{k}}{\left( 2\pi\right)^3 }\, \frac{d\mathbf{q}}{\left( 2\pi\right)^3 } \,e^{i\,\mathbf{k}\,\mathbf{r}} \,e^{i\,\mathbf{q}\,\mathbf{v}} }\\ %\\
{\displaystyle \times \iint d\mathbf{R}\, d\mathbf{V}\, f\left(\mathbf{R}, \mathbf{V}, t\right)\, e^{-i\, \mathbf{k R}}\,  e^{-i\, \mathbf{q V}} \, \left(-i\, \mathbf{k V} \right) }\\
{\displaystyle = \int d\mathbf{R}\,  f\left(\mathbf{R}, \mathbf{v}, t\right)\, \left\lbrace \int \frac{d \mathbf{k}}{\left(2\pi  \right)^3 }\ e^{i\, \mathbf{k \left(r - R \right) }}\, \left(-i\, \mathbf{k v} \right)    \right\rbrace }.
\end{array}
\end{equation}

Using the identity 
\begin{equation}\label{J-full}
J = \int \frac{d \mathbf{k}}{\left(2\pi  \right)^3 }\ e^{i\, \mathbf{k r  }}\, \left(\mathbf{k v} \right) =  -i \left(\mathbf{v} \cdot \nabla \right)\, \delta\left( \mathbf{r}\right) 
\end{equation}
to this formula leads to the final expressions for $\left( {\partial f(\mathbf{r},\mathbf{v},t)}/{\partial t}\right) _{1} $:
\begin{equation}\label{dot-f1-1}
\begin{array}{c}
{\displaystyle \left( \frac{\partial f(\mathbf{r},\mathbf{v},t)}{\partial t}\right) _{1} =  - \frac{\partial }{\partial \mathbf{r}}\left(\mathbf{v}\, f\left(\mathbf{r}, \mathbf{v}, t\right)  \right)}\\
{\displaystyle  = -\left( \mathbf{v}\cdot \frac{\partial }{\partial \mathbf{r}} \right)  f\left(\mathbf{r}, \mathbf{v}, t\right).}
\end{array}
\end{equation}
%

%\subsection{Calculation of  $\left( \frac{\partial f(\mathbf{r},\mathbf{v},t)}{\partial t}\right) _{2}$}
%
To calculate the function $\left( \frac{\partial f(\mathbf{r},\mathbf{v},t)}{\partial t}\right) _{2}$, it is necessary to express $ \ddot{\mathbf{R}}_s(t) $ via the microscopic distribution function~$f(\mathbf{r},\mathbf{v},t)$. According to Newton's second law
\begin{equation}\label{Newton}
	\ddot{\mathbf{R}}_s \left(t \right) = - \frac{1}{m}\, \nabla \, \varPhi\left( \mathbf{R}_s, t\right),
\end{equation}
where $ \varPhi\left( \mathbf{R}_s, t\right)$~ is the field's potential acting on the $s$-th particle by all the particles
\begin{equation}\label{loc-field}
	\begin{array}{c}
		{\displaystyle \varPhi\left( \mathbf{R}_s, t\right) = \sum_{s'}\ W\left(\mathbf{R}_s - \mathbf{R}_{s'}\left( t - \tau\right) \right) }\\
		{\displaystyle = \iint d\mathbf{R}'\, d\mathbf{V}' \,  W\left(\mathbf{R}_s - \mathbf{R}'\right)\, f(\mathbf{R}',\mathbf{V}',t - \tau \left(\mathbf{R}_s - \mathbf{R}' \right) )}, 
	\end{array}
\end{equation}
$W\left(\mathbf{R}_s - \mathbf{R}_{s'}\right) $ is the potential energy of interaction between two \textit {resting} particles, situated at the points $\mathbf{R}_s $ and $ \mathbf{R}_{s'} $, $\tau \left(\mathbf{R} - \mathbf{R}' \right)$ is the interactions retardation between these points.

Substitute~(\ref{Newton}) and~(\ref{loc-field}) into (\ref{dot2-f}) and find
\begin{equation}\label{dot-f2-1}
	\begin{array}{c}
		{\displaystyle \left( \frac{\partial f(\mathbf{r},\mathbf{v},t)}{\partial t}\right) _{2} = \frac{i}{m} \int \frac{d\mathbf{q}}{\left( 2\pi\right)^3 }\, e^{i\, \mathbf{q v}}\, \int  d\mathbf{V}'\, e^{-i\,\mathbf{q}\,\mathbf{V}'}\, f\left( \mathbf{r}, \mathbf{V}', t \right) } \\ \\
		{\displaystyle \times \, \left(\mathbf{q}\cdot \frac{\partial}{\partial \mathbf{r}} \iint d\mathbf{R}\, d\mathbf{V}\, f\left( \mathbf{R}, \mathbf{V}, t - \tau \left( \left| \mathbf{r} - \mathbf{R} \right|  \right)  \right) W\left(\mathbf{r} - \mathbf{R} \right)  \right).}
	\end{array}
\end{equation}
Using the identity~(\ref{J-full}) leads to a significant simplification of this formula
\begin{equation}\label{dot-f2-2}
\begin{array}{c}
{\displaystyle \left( \frac{\partial f(\mathbf{r},\mathbf{v},t)}{\partial t}\right) _{2} }  \\ 	\\	{\displaystyle 
	= \frac{1}{m} \biggl( \frac{\partial f\left(\mathbf{r}, \mathbf{v}, t \right) }{\partial \mathbf{v}}  \cdot \frac{\partial}{\partial \mathbf{r}} \iint  f\left( \mathbf{R}, \mathbf{V}, t - \tau \left( \left| \mathbf{r} - \mathbf{R} \right|  \right)  \right)}\\ {\displaystyle \times\ W\left(\mathbf{r} - \mathbf{R} \right)\, d\mathbf{R}\, d\mathbf{V} \biggr).}
\end{array}
\end{equation} 
\subsection{Basic equation}

Combining relations~(\ref{dot1+2-f}), (\ref{dot-f1-1}) and (\ref{dot-f2-2}), we find \textit{the exact closed} basic equation for the microscopic distribution function~$f(\mathbf{r},\mathbf{v},t)$:  
\begin{widetext}
	\begin{equation}\label{main-eq}
 \left( \frac{\partial f(\mathbf{r},\mathbf{v},t)}{\partial t}\right)\, + \,\left( \mathbf{v}\cdot \frac{\partial f\left(\mathbf{r}, \mathbf{v}, t \right) }{\partial \mathbf{r}} \right) = \, \frac{1}{m} \biggl( \frac{\partial f\left(\mathbf{r}, \mathbf{v}, t \right) }{\partial \mathbf{v}}  \cdot \frac{\partial}{\partial \mathbf{r}} \iint f\left( \mathbf{R}, \mathbf{V}, t - \tau \left( \left| \mathbf{r} - \mathbf{R} \right|  \right)  \right) \ W\left(\mathbf{r} - \mathbf{R} \right)\, d\mathbf{R}\, d\mathbf{V}  \biggr). 
	\end{equation}
In the presence of an external field $\varphi\left(\mathbf{r}, t \right)  $, this equation has the form
\begin{equation}\label{main-equat-ext}
\begin{array}{c}
{\displaystyle \left( \frac{\partial f(\mathbf{r},\mathbf{v},t)}{\partial t}\right)\, + \,\left( \mathbf{v}\cdot \frac{\partial f\left(\mathbf{r}, \mathbf{v}, t \right) }{\partial \mathbf{r}} \right) - \frac{1}{m} \biggl( \frac{\partial f\left(\mathbf{r}, \mathbf{v}, t \right) }{\partial \mathbf{v}}  \cdot \frac{\partial \varphi\left(\mathbf{r}, t \right) }{\partial \mathbf{r}}   \biggr) }\\
{\displaystyle   = \, \frac{1}{m} \biggl( \frac{\partial f\left(\mathbf{r}, \mathbf{v}, t \right) }{\partial \mathbf{v}}  \cdot \frac{\partial}{\partial \mathbf{r}} \biggl[ \iint f\left( \mathbf{R}, \mathbf{V}, t - \tau \left( \left| \mathbf{r} - \mathbf{R} \right|  \right)  \right) \ W\left(\mathbf{r} - \mathbf{R} \right) d\mathbf{R}\, d\mathbf{V} \biggr]  \biggr).}
\end{array}
\end{equation}

\end{widetext}

\section{Irreversibility of the basic equation}
Let us consider  the qualitative properties of the basic equation~(\ref{main-eq}), with special emphasis on the time-irreversibility problem of its solutions.  For this purpose we expand the integrand in the basic equation~(\ref{main-eq}) in powers of~$ \tau \left( \left| \mathbf{r} - \mathbf{R} \right|  \right) $ and substitute the result in this equation. As a result we have
\begin{widetext}
\begin{equation}\label{righr-hand}
\begin{array}{c}
{\displaystyle \left( \frac{\partial f(\mathbf{r},\mathbf{v},t)}{\partial t}\right)\, + \,\left( \mathbf{v}\cdot \frac{\partial f\left(\mathbf{r}, \mathbf{v}, t \right) }{\partial \mathbf{r}} \right) }\\ 
{\displaystyle = \frac{1}{m} \sum_{n=0}^{\infty}(-1)^{n} \biggl( \frac{\partial f\left(\mathbf{r}, \mathbf{v}, t \right) }{\partial \mathbf{v}}  \cdot \frac{\partial}{\partial \mathbf{r}} \iint \left\lbrace W\left(\mathbf{r} - \mathbf{R} \right) \left[ \tau\left(\left|\mathbf{r}-\mathbf{R} \right|  \right)\right] ^{n}\right\rbrace \, \left[ \left( \dfrac{\partial } {\partial t} \right)^{n}  f\left( \mathbf{R}, \mathbf{V}, t \right)\right]  \, d\mathbf{R}\, d\mathbf{V}  \biggr)}.
\end{array}
\end{equation}	
\end{widetext}

Let us consider the changes every of the terms in this equation under the Loschmidt's  time reversal transformation
\begin{equation}\label{time-rev}
t \rightarrow -t,\ \mathbf{v} \rightarrow -\mathbf{v},\ \mathbf{V} \rightarrow -\mathbf{V}.
\end{equation}
According to the definition~(\ref{f(r,v,t)}), we have
\begin{eqnarray}\label{Loschmidt}
 f(\mathbf{r},\mathbf{v},t) \rightarrow f(\mathbf{r},-\mathbf{v},-t) = f(\mathbf{r},\mathbf{v},t);\label{f(t)} \\
  \frac{\partial f(\mathbf{r},\mathbf{v},t) }{\partial t} \rightarrow \frac{\partial f(\mathbf{r},-\mathbf{v},-t)}{\partial  t}  = - \frac{\partial f(\mathbf{r},\mathbf{v},t) }{\partial  t} ;\label{f2(t)} \\
  \frac{\partial f(\mathbf{r},\mathbf{v},t) }{\partial \mathbf{r}} \rightarrow \frac{\partial f(\mathbf{r},-\mathbf{v},-t)}{\partial  \mathbf{r} }  =  \frac{\partial f(\mathbf{r},\mathbf{v},t) }{\partial  \mathbf{r}} ; \\
  \frac{\partial f(\mathbf{r},\mathbf{v},t) }{\partial \mathbf{v}} \rightarrow \frac{\partial f(\mathbf{r},-\mathbf{v},-t)}{\partial  \mathbf{v} }  = - \frac{\partial f(\mathbf{r},\mathbf{v},t) }{\partial  \mathbf{v}} ;\\
 \left( \dfrac{\partial } {\partial t} \right)^{n}  f\left( \mathbf{R}, \mathbf{V}, t \right) \rightarrow  \left( \dfrac{\partial } {\partial t} \right)^{n}  f\left( \mathbf{R}, -\mathbf{V}, -t \right)\\  \nonumber = (-1)^{n} \left( \dfrac{\partial } {\partial t} \right)^{n}  f\left( \mathbf{R}, \mathbf{V}, t \right).
\end{eqnarray}
Thus, the time reversal operation leads to change of left-hand side sign of equation~(\ref{righr-hand}). In the right-hand side of this equation a such change of sign takes a place for the terms with even values~$n$ only.

Therefore, equation~(\ref{righr-hand}) is invariant with respect to time reversal if and only if there is no delay interactions, i.e. $\tau(\mathbf{r} ) \equiv 0$.
In other words, the interactions delay leads to irreversible behavior of a system of interacting particles. 
In this case, the equation~(\ref{main-eq}) is reversible and has a well-known form~\cite{Kadomtsev}:
	\begin{equation}\label{main-eq2}
	\begin{array}{c}
	{\displaystyle  \left( \frac{\partial f(\mathbf{r},\mathbf{v},t)}{\partial t}\right)\, + \,\left( \mathbf{v}\cdot \frac{\partial f\left(\mathbf{r}, \mathbf{v}, t \right) }{\partial \mathbf{r}} \right) = \, \frac{1}{m}}\\
	{\displaystyle \times \biggl( \frac{\partial f\left(\mathbf{r}, \mathbf{v}, t \right) }{\partial \mathbf{v}}  \cdot \frac{\partial}{\partial \mathbf{r}} \iint f\left( \mathbf{R}, \mathbf{V}, t  \right) \ W\left(\mathbf{r} - \mathbf{R} \right)\, d\mathbf{R}\, d\mathbf{V}  \biggr). }
	\end{array}
	\end{equation}
Note that this equation for the microscopic (not probabilistic)  distribution function is reversible with respect to time reversal.

\section{Physical interpretation of first terms expansion over interactions retardation}

Let us consider the expansion~(\ref{main-equat-ext}) in powers of the interactions retardation up to the first non-vanishing term
\begin{widetext}
\begin{eqnarray}
{\displaystyle \left( \frac{\partial f(\mathbf{r},\mathbf{v},t)}{\partial t}\right)\, + \,\left( \mathbf{v}\cdot \frac{\partial f\left(\mathbf{r}, \mathbf{v}, t \right) }{\partial \mathbf{r}} \right) - \frac{1}{m} \biggl( \frac{\partial f\left(\mathbf{r}, \mathbf{v}, t \right) }{\partial \mathbf{v}}  \cdot \frac{\partial \varphi\left(\mathbf{r}, t \right) }{\partial \mathbf{r}}   \biggr) }\label{left-Boltz}\\
{\displaystyle =  \frac{1}{m} \biggl( \frac{\partial f\left(\mathbf{r}, \mathbf{v}, t \right) }{\partial \mathbf{v}}  \cdot \frac{\partial}{\partial \mathbf{r}} \biggl[ \iint f\left( \mathbf{R}, \mathbf{V}, t \right) \ W\left(\mathbf{r} - \mathbf{R} \right) d\mathbf{R}\, d\mathbf{V} \biggr]  \biggr) }\label{Vlasov}\\
{\displaystyle  +  \frac{1}{m} \biggl( \frac{\partial f\left(\mathbf{r}, \mathbf{v}, t \right) }{\partial \mathbf{v}}  \cdot \frac{\partial}{\partial \mathbf{r}} \iint \left\lbrace W\left(\mathbf{r} - \mathbf{R} \right)\cdot   \tau\left(\left|\mathbf{r}-\mathbf{R} \right|  \right) \right\rbrace \,  \left( \dfrac{\partial  f\left( \mathbf{R}, \mathbf{V}, t \right)} {\partial t} \right)  \, d\mathbf{R}\, d\mathbf{V}  \biggr).\label{retard1}}
\end{eqnarray}
\end{widetext}
Expression~(\ref{left-Boltz}) corresponds completely by the form  to the left-hand side of the Boltzmann equation, taking into account the external field~$\varphi\left(\mathbf{r}, t \right)$. The second term~(\ref{Vlasov}) is similar in its form to the self-consistent field term in the Vlasov equation. However, the corresponding term in the Vlasov equation contains  \textit{probabilistic distribution function} and therefore it is some \textit{approximate} expression. Equation~(\ref{righr-hand}) contains the microscopic \textit{non-averaged} distribution function and therefore it is the \textit{exact} expression. The third term~(\ref{retard1}) has a remote likeness with collision integral on the retarded part of the field generated by all the particles of the system. In this case, instead of the interatomic potential~$W(\mathbf{r})$ this part of the equation contains the ``renormalized potential'' of the form~$W(\mathbf{r})\, \tau(\mathbf{r})$.

\section{Discussion of the results}

The exact equation of motion for the microscopic distribution function of a system consisting of identical interacting particles is derived. The interaction potential between the particles assumed an arbitrary function of the distance. Effect of the interactions retardation is taken into account.

Despite the external resemblance of the basic equation~(\ref{main-eq}) with the Boltzmann equation, there are fundamental differences between these equations.
\begin{enumerate}
	\item The Boltzmann equation contains a \textit{ probabilistic } distribution function, while the basic equation~(\ref{main-eq}) contains the exact \textit{ microscopic } distribution function.
	\item The Boltzmann equation is based on a simplified conception that the  particles are moving freely between their collisions. Interactions between particles are taken into account by the collision integral, having a probabilistic character. This picture is applicable (to some extent) to the rarefied gas, consisting of hard spheres. Unlike Boltzmann equation, the basic equation~(\ref{main-eq}) is free from all these limitations. The interaction between the particles is described by an \textit{arbitrary} potential~$W(\mathbf{r})$. Each of the particles is influenced by all the other particles.
	\item The interaction between the particles in the framework of the Boltzmann equation is taken into account by the collision integral. In this connection time evolution of the \textit{probabilistic} distribution function is represented as a sequence of mainly pairwise collisions of particles. In the framework of the basic \textit{deterministic} equation~(\ref{main-eq}), each particle interacts continuously with all the other particles without separation on two-body, three-body, four-body, \ldots\ collisions.
\end{enumerate}

Certainly, apart from the Boltzmann equation, there are other methods to describe the kinetic processes~\cite{Keizer,Gallavotti}. However, all existing methods of kinetic processes description, including BBGKY approach~\cite{Bog,Born,Kirkwood,Yvon}, are based on probabilistic  hypotheses and assumptions.

\section{Conclusion}

The statistical approach to the theory of many-particle systems originates since the late 19th century. This approach is the basis of equilibrium and non-equilibrium statistical mechanics having enormous success in the quantitative description of the properties of many-particle systems. However, there are a number of fundamental problems in the foundations of statistical mechanics and in the understanding of its essence. 

Note the fundamental difference between problems settings in the mechanics of many-particle systems with instantaneous interactions and with retarded interactions. In particular, it should be noted, that the standard setting of the Cauchy problem for with retarted interactions is not sufficient for uniqueness of these equations solution. Namely, a solution of equation~(\ref{main-eq}) depends not only on the initial conditions, but also on the previous history of the system.
Thus, the inclusion of the interactions delay leads to drastic change in the qualitative properties of the equation~(\ref{main-eq}) solutions.

I am grateful to Prof. Ya.I. Granovsky and Prof.~Istv\'an Mayer for numerous discussions and helpful comments on this paper.

This work was partially supported by the Russian Ministry of Science and Education in the framework of the base part of state order (Project number 1755).

%\vskip1cm

\end{document}